%% file: main.tex
\documentclass[reprint,secnumarabic,amssymb, nobibnotes, aps, prd,floatfix]{revtex4-1}
\usepackage{amsmath}
\usepackage{graphicx}
\usepackage{hyperref}

\usepackage[dvipsnames]{xcolor}

\newcommand{\bs}{\boldsymbol}
\newcommand{\Tr}{\mathrm{Tr}}

\newcounter{myequation}
\makeatletter
\@addtoreset{equation}{myequation}
\makeatother
\newcounter{myfigure}
\makeatletter
\@addtoreset{figure}{myfigure}
\makeatother

\begin{document}
\begin{abstract}
Turbulent fluid flows exhibit a complex small-scale structure with frequently occurring extreme velocity gradients. Particles probing such swirling and straining regions respond with an intricate shape-dependent orientational dynamics, which sensitively depends on the particle history.  Here, we systematically develop a reduced-order model for the small-scale dynamics of turbulence, which captures the velocity gradient statistics along particle paths. An analysis of the resulting stochastic dynamical system allows pinpointing the emergence of non-Gaussian statistics and nontrivial temporal correlations of vorticity and strain, as previously reported from experiments and simulations. Based on these insights, we use our model to predict the orientational statistics of anisotropic particles in turbulence, enabling a host of modeling applications for complex particulate flows.
\\
\\
\noindent
DOI: \href{https://doi.org/10.1103/PhysRevLett.125.224501}{https://doi.org/10.1103/PhysRevLett.125.224501}
\end{abstract}

\title{Capturing Velocity Gradients and Particle Rotation Rates in Turbulence}

\author{Leonhard A. Leppin$^{1,2,3}$}
\author{Michael Wilczek$^{1,2}$}
\email[Michael Wilczek: ]{michael.wilczek@ds.mpg.de}
\affiliation{${}^{1}$Max Planck Institute for Dynamics and Self-Organization (MPI DS), Am Fa\ss berg 17, 37077 G\"ottingen, Germany \\
\mbox{${}^{2}$Faculty of Physics, Georg-August-Universit\"at G\"ottingen, Friedrich-Hund-Platz 1, 37077 G\"ottingen, Germany}
\\
${}^{3}$Max Planck Institute for Plasma Physics, Boltzmannstra\ss e 2, 85748 Garching, Germany}
\maketitle

Turbulent flows show complex dynamics with a wide range of dynamically active scales \cite{pope_turbulent_2000,frisch_turbulence:_1995,tsinober_informal_2009}, which play an important role for the dispersal of pollutants and aerosols in the atmosphere \cite{fernando_flow_2010,chun_clustering_2005}, the transport of microorganisms in the ocean \cite{durham_turbulence_2013,abraham_generation_1998,de_lillo_turbulent_2014,gustavsson_preferential_2016,breier_emergence_2018}, as well as the mixing of reactants in turbulent combustion \cite{peters_turbulent_2000,bilger_turbulent_1989}. The smallest turbulent scales, which are essentially independent of the boundaries and anisotropies of the large-scale flow \cite{pope_turbulent_2000}, have a profound impact on the dynamics and collision rates of small suspended particles, like plankton in the ocean \cite{durham_turbulence_2013,abraham_generation_1998,de_lillo_turbulent_2014,gustavsson_preferential_2016,breier_emergence_2018} as well as droplets and ice crystals in clouds \cite{falkovich_acceleration_2002,shaw_particle-turbulence_2003,jucha_settling_2018,bodenschatz_can_2010}. Even in the simplest case of very small, neutrally buoyant particles, which passively follow the velocity field, highly nontrivial, shape-dependent rotational motion has been observed \cite{voth_anisotropic_2017,chevillard_orientation_2013,parsa_rotation_2012,byron_shape-dependence_2015,challabotla_orientation_2015,ni_measurements_2015}. Theoretically, this intricate dynamics is not well understood.

The spinning and tumbling of particles immersed in a turbulent flow are determined by the complex interplay of particle shape and the small-scale structure of the turbulent flow field, as encoded in the gradients of the velocity field $A_{ij}=\partial u_i/\partial x_j$ \cite{meneveau_lagrangian_2011}. Because particle rotations are very sensitive to various small-scale features of turbulence such as non-Gaussian fluctuations, the local flow topology and, most importantly, the temporal correlation of strain and vorticity along Lagrangian trajectories, capturing this complex motion with theoretically insightful reduced-order models for turbulence so far remained elusive. The challenges in predicting these aspects of turbulent velocity gradients ultimately arise from the nonlinear, nonlocal, and dissipative dynamics of the governing Navier-Stokes equations.

Over the past years, a variety of reduced-order models for the velocity gradient statistics based on stochastic differential equations (SDEs) has been developed \cite{meneveau_lagrangian_2011,girimaji_diffusion_1990,chertkov_lagrangian_1999,naso_scale_2005,chevillard_lagrangian_2006,wilczek_pressure_2014,johnson_closure_2016}. In these models, the effects of nonlocal pressure and viscous diffusion result in unclosed terms, to which diverse closure techniques have been applied. Closure theories range from models based on prescribed log-normal dissipation rates \cite{girimaji_diffusion_1990}, the coarse-grained velocity gradient as perceived by a tetrad of tracer particles \cite{chertkov_lagrangian_1999,naso_scale_2005}, or the deformation of fluid elements \cite{chevillard_lagrangian_2006} to functional closures based on Gaussian random fields \cite{wilczek_pressure_2014}, as well as combinations of these approaches \cite{johnson_closure_2016}. The most advanced reduced-order SDE models successfully reproduce many of the characteristic geometric and statistical properties of the turbulent small scales \cite{chevillard_modeling_2008,wilczek_pressure_2014,johnson_closure_2016,johnson_turbulence_2017,johnson_predicting_2018}. However, all current models struggle to capture important aspects of the temporal correlation of strain rate and rotation rate, which, in particular, leads to poor predictions for the orientational dynamics of particles immersed in turbulent flows.

Here, we develop a minimal model for the velocity gradients in turbulence, which enables profound theoretical insights. Starting from an exact statistical evolution equation, we systematically constrain its structure based on tensor function representation theory. By using an ensemble approach, we construct a physically consistent model which complies with important homogeneity constraints of turbulent fields. Based on an analysis of the associated Fokker-Planck equation, we establish a clear interpretation of its nonlinear dynamics. Specifically, we identify the dynamical mechanisms which control the degree of non-Gaussianity and temporal correlations of vorticity and strain. We test our predictions against high-resolution simulation results of fully developed turbulence and show that our model captures the temporal autocorrelations of rotation rate and strain rate. Coupled to the equations for the orientation dynamics of ellipsoidal particles, our model, furthermore, accurately reproduces the tumbling and spinning rates of particles in turbulent flows.

The evolution equation for velocity gradients is obtained by taking the gradient of the Navier-Stokes equation. Along a tracer particle, the velocity gradient changes according to the local self-amplification of velocity gradients, nonlocal pressure contributions, viscous diffusion, and external forces \cite{meneveau_lagrangian_2011}. The foundation of our model is an exact, unclosed SDE for the one-point statistics of homogeneous isotropic turbulence, which statistically captures these various contributions to the dynamics.
Using stochastic calculus, one can derive this SDE from the Navier-Stokes equations \cite{wilczek_pressure_2014}. It takes the form
\begin{equation}
\label{eq:sde}
d\bs{A}=\left(-\widetilde{\bs{A}^2}-\langle\widetilde{\bs{H}}|\bs{A}\rangle+\langle\nu\Delta \bs{A}|\bs{A} \rangle\right)dt + d\bs{F} \,.
\end{equation}

Here, $\bs{A}$ can be interpreted as the stochastic process corresponding to the velocity gradient field at the position of a fluid particle. The tilde denotes the traceless part of the tensor, e.g.~$\widetilde{\bs{A}^2}=\left(\bs{A}^2-\frac{1}{3}\Tr(\bs{A}^2) \bs I \right)$.  The first term on the right-hand side, which appears in closed form, captures the nonlinear local self-amplification of the velocity gradient. It includes the local isotropic part of the pressure Hessian $H_{ij}=\frac{\partial^2 p}{\partial x_i \partial x_j}$, which is obtained from the pressure Poisson equation $\Delta p = \mathrm{Tr}(\bs H) = - \mathrm{Tr}(\bs A^2)$ \cite{ohkitani_nonlocal_1995}. Unclosed terms, which contain information beyond the local single-point statistics, appear in the form of conditional averages, i.e.,~as averaged fields conditional on a given configuration of the velocity gradient at the same position. The conditional average $\langle\widetilde{\bs{H}}|\bs{A}\rangle$ contains information about the mean nonlocal, deviatoric part of the pressure Hessian given a velocity gradient configuration $\bs{A}$, which a priori depends on the full flow field due to the pressure Poisson equation. The conditional Laplacian $\langle\nu\Delta \bs{A}|\bs{A} \rangle$ encodes viscous effects in the velocity gradient tensor evolution. The term $d\bs{F}$ is a Gaussian, temporally delta-correlated tensorial forcing, which is consistent with isotropy, homogeneity and incompressibility $\Tr(\bs{A})=0$. It naturally arises when considering stochastically forced turbulence.

\begin{figure*}
	\includegraphics[width=\linewidth]{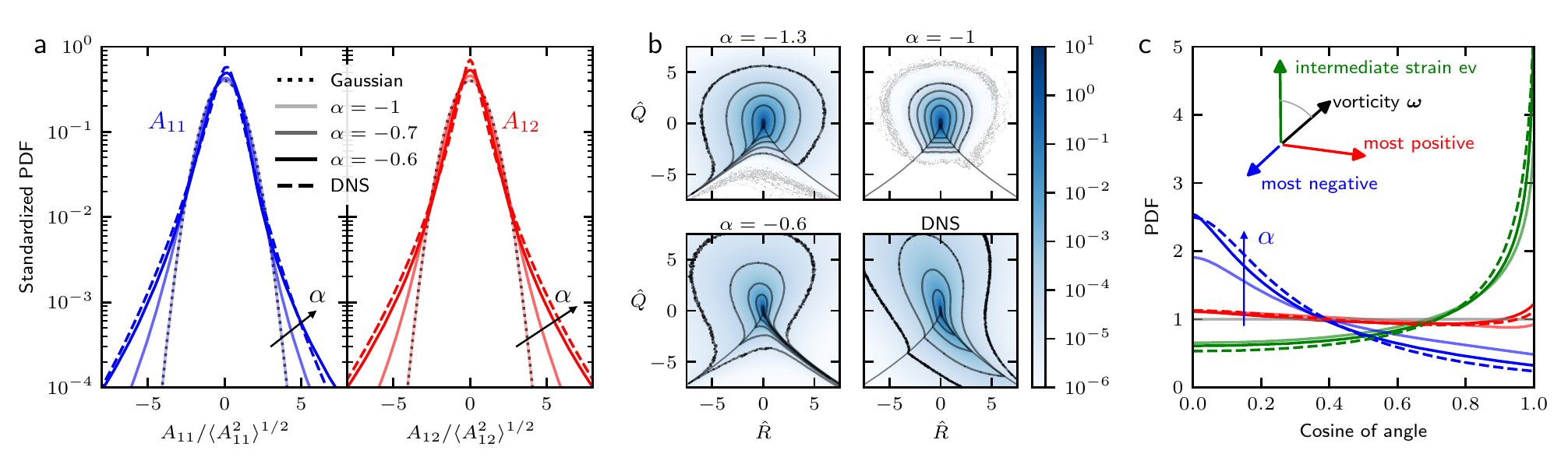}
	\caption{Controlling single-time statistics with the strength of the strain self-amplification $\alpha$. (a) $\alpha$ controls the departure from Gaussianity of the model statistics: Standardized PDFs of the longitudinal ($A_{11}$) and transverse ($A_{12}$) velocity gradient components for different values of $\alpha$. (b) $\alpha$ determines the probability of different flow topologies: Joint PDF of the standardized isotropic invariants $\hat{R}=-\frac{1}{3}\frac{\Tr(\bs{A}^3)}{\langle\Tr(\bs{S}^2) \rangle^{3/2}}$ and $\hat{Q}=-\frac{1}{2}\frac{\Tr(\bs{A}^2)}{\langle\Tr(\bs{S}^2)\rangle}$ for different values of $\alpha$. DNS results in lower right panel. The Vieillefosse line is indicated in gray. (c) Alignment of principal strain axes and vorticity increases with strain self-amplification: PDFs of the cosine of the angle between the vorticity vector and the three eigenvectors of the strain-rate tensor for the same values of $\alpha$ as in (a).}
	\label{fig:fig1}
\end{figure*}

To close the conditional mean pressure Hessian and the conditional mean Laplacian terms, we express them as isotropic tensor-valued functions of the symmetric and antisymmetric part of the velocity gradient, the strain rate $\bs{S}$, and rotation rate $\bs{W}$, respectively. Using tensor function representation theory, one can derive a complete and irreducible representation in terms of a small number of tensorial terms \cite{wang_new_1970,smith_isotropic_1971,boehler_irreducible_1977,pennisi_irreducibility_1987,zheng_theory_1994,boehler_applications_1987}. The individual tensorial terms are comprised of combinations of $\bs{S}$ and $\bs{W}$, with coefficient functions that depend on isotropic invariants of $\bs{S}$ and $\bs{W}$. The conditional mean traceless, symmetric pressure Hessian, for example, can be expressed as a linear combination of seven tensorial terms with appropriate coefficient functions (cf. Supplemental Materials \cite{sm}, Sec. \ref{sec:sm1}). Previous studies \cite{wilczek_pressure_2014,lawson_velocity_2015} showed that the most important features of the dynamics can already be captured by retaining terms up to the lowest possible order (i.e.,~up to second order in the pressure Hessian and up to first order in the viscous term) with constant coefficients. To enable analytical insights, we therefore truncate the general tensorial expansion and consider the closure
\begin{align}
\langle\widetilde{\bs{H}}|\bs{A}\rangle&=\alpha\widetilde{\bs{S}^2}+\beta\widetilde{\bs{W}^2}+\gamma(\bs{SW}-\bs{WS})+\delta\bs{S} \label{eq:pressurehessianclosure} \\
\langle\nu \Delta \bs{A}|\bs{A}\rangle&=\xi\bs{A} \label{eq:laplacianclosure} \, .
\end{align}
This expression still contains five scalar parameters, which need to be further constrained. A general limitation of single-point closures is that they lack the possibility to include physical constraints that depend on information from the full field. For homogeneous turbulence, for example, the velocity gradient field fulfills the Betchov constraints \cite{betchov_inequality_1956} $\langle \Tr(\bs{A}^2)\rangle=0$ and $\langle \Tr(\bs{A}^3)\rangle=0$, which encode the balance of enstrophy and dissipation, as well as of their production. So far, velocity gradient models need careful calibration to fulfill these constraints \cite{johnson_closure_2016}. An intriguing alternative to achieve a model that is physically consistent with homogeneous turbulence is to consider an ensemble of Lagrangian fluid elements that sample the full velocity gradient field. We then achieve consistency with the Betchov constraints by identifying spatial averages over the field with ensemble averages over the Lagrangian fluid elements. This can be used to derive analytical expressions from \eqref{eq:sde} for two of the parameters, which allows us to constrain our closure \eqref{eq:pressurehessianclosure} and \eqref{eq:laplacianclosure}. One additional parameter can be fixed by nondimensionalizing the velocity gradient model with the Kolmogorov time scale $\tau_\eta$, which implies $\langle\Tr(\bs{S}^2)\rangle=1/2$. We choose to constrain the parameters $\beta$, $\delta$, and $\xi$ and obtain them as functions of ensemble-averaged scalar invariants of the velocity gradient and the remaining parameters. Their explicit form and derivation is given in Supplemental Material, Sec. \ref{sec:sm2} \cite{sm}. Thereby, the parameter space is reduced by three dimensions, and Betchov's homogeneity constraints are fulfilled by design. Besides the forcing amplitude, which we fix for the following considerations (see Supplemental Material \cite{sm}, Sec. \ref{sec:forcing} for the impact of the forcing amplitude), this leaves two free parameters: $\alpha$ and $\gamma$.

The impact of these free parameters on the nonlinear dynamics of the velocity gradient model can be revealed from the Fokker-Planck equation (FPE) corresponding to \eqref{eq:sde}, which governs the evolution of the full probability density function (PDF) $f(\bs{A};t)$ of the velocity gradient tensor (implied summation),
\begin{equation}
\label{eq:fokplacomp}
\frac{\partial}{\partial t}f = -\frac{\partial}{\partial A_{ij}}\left[ (N_{ij}+L_{ij})f \right]+\frac{1}{2}Q_{ijkl}(\bs{0})\frac{\partial}{\partial A_{ik}}\frac{\partial}{\partial A_{jl}}f \, .
\end{equation}
Here, $Q_{ijkl}(\bs{0})$ denotes the forcing covariance, and the nonlinear and linear drift terms are given by
\begin{align}
\bs{N}=&-(1+\alpha) \widetilde{\bs{S}^2}-(1+\beta) \widetilde{\bs{W}^2} \nonumber\\
			 &-(1+\gamma) \bs{SW}-(1-\gamma)\bs{WS}\\
\bs{L}=&-\delta \bs{S}+\xi \bs{A} \, .
\end{align}
The parameter $\alpha$ controls the strength of the strain self-amplification in the velocity gradient dynamics. For a vanishing self-amplification ($\alpha=-1$) and parameters determined such that the Betchov constraints are fulfilled, we find that \eqref{eq:fokplacomp} has an exact Gaussian solution (cf. Supplemental Material \cite{sm}, Sec. \ref{sec:sm41}). Remarkably, even in this case, the FPE contains nonlinear drift terms.  We demonstrate below that the strength of strain self-amplification controls departures from Gaussianity as well as important features of the small-scale topology of the predicted velocity gradient statistics. Further analysis of the FPE shows that the single-time statistics is independent of the parameter $\gamma$ for isotropic turbulence. In this case, the velocity gradient PDF is a function of the tensor invariants only, and one can readily calculate that $\frac{\partial}{\partial A_{ij}}\gamma[(S_{ik}W_{kj}-W_{ik}S_{kj})f(\bs{A})]=0$ (cf. Supplemental Material \cite{sm}, Sec. \ref{sec:sm42}). This result is related to a recently reported gauge symmetry of the pressure Hessian \cite{carbone_gauge_2020}. However, we show below that, for the two-time statistics, and, in particular the autocorrelations of vorticity and strain, the $\gamma$-term turns out to be crucial.

To determine appropriate values for the free parameters, we perform parameter scans and compare our model results with velocity gradient statistics obtained from direct numerical simulations (DNS) of the Navier-Stokes equation. We analyzed a simulation of homogeneous isotropic turbulence with $2048^3$ grid points on a periodic domain at a Taylor-scale Reynolds number of $R_\lambda \approx 509$ with a small-scale resolution of $k_{\mathrm{max}}\eta\approx1.5$ ($k_{\mathrm{max}}$ is the largest resolved wavenumber, and $\eta$ is the Kolmogorov scale). For the data analysis, 25 snapshots spanning ca.~four integral time scales are taken into account. For the parameter scans, we numerically solve \eqref{eq:sde} using the Euler-Maruyama method \cite{kloeden_numerical_1999} with a time step of $\Delta t = 0.0002$. Convergence checks with varying time steps were also performed. For all simulations shown here, we have integrated an ensemble of $10^5$ Gaussian initial conditions for $5 \times 10^6$ time steps, which corresponds to $1000\tau_\eta$ in physical time, after an initial transient of $100\tau_\eta$ (cf. Supplemental Material \cite{sm}, Sec. \ref{sec:sm5}). Initial simulations of \eqref{eq:sde} revealed the occurrence of rare rogue trajectories exploring far-out regions of the phase space, which leads to nonconvergent statistics and may introduce numerical instabilities in the determination of our parameters. We identified the second-order truncation of the unclosed terms as the origin of this shortcoming, which can be remedied by including a nonlinear damping term (cf. Supplemental Material \cite{sm}, Sec. \ref{sec:dampingterm}). The auxiliary term  $\epsilon \bs A$, which is added to \eqref{eq:laplacianclosure}, is constructed to damp trajectories that diverge far from the ensemble mean and is negligibly small for the major, dynamically most relevant part of phase space; we set $\epsilon=-10^{-8}\{[\Tr(\bs{W}^2)+1/2]^4+[\Tr(\bs{S}^2)-1/2]^4\}$. 

Figure \ref{fig:fig1}(a) illustrates how the strength of strain self-amplification controls the departure from Gaussianity of the predicted velocity gradient statistics.
As $\alpha$ deviates from $-1$, the single-component PDFs become non-Gaussian with increasingly heavy tails.
For $\alpha=-0.6$, the standardized PDFs of the velocity gradient components of our model agree well with DNS results within seven standard deviations [cf. Fig.~\ref{fig:fig1}(a)]. The match is particularly good for the transverse components, for which our model captures the vanishing skewness and closely matches the DNS kurtosis of  $\langle A_{12}^4 \rangle / \langle A_{12}^2 \rangle^2 \approx 14.34$ to within 1\% (model: 14.32). For the longitudinal components, our model underpredicts the skewness $\langle A_{11}^3 \rangle / \langle A_{11}^2 \rangle^{3/2}$ (Model: -0.42, DNS: -0.61), consistent with other recent models \cite{johnson_closure_2016}, and slightly overpredicts the kurtosis (Model: 11.46, DNS: 9.2).

Furthermore, the strain self-amplification determines the probability of different flow topologies, as encoded in the invariants $Q=-\Tr(\bs{A}^2)/2$ and $R=-\Tr(\bs{A}^3)/3$, which capture the competition between enstrophy and dissipation, as well as their production. In the $R$-$Q$ plane, the Vieillefosse line, i.e.,~the zero crossing of the discriminant of $\bs A$ given by $(27/4)R^2+Q^3=0$, plays an important role, as it separates the upper region with complex eigenvalues of $\bs A$ from the lower region with purely real eigenvalues. Figure \ref{fig:fig1}(b) shows the joint PDF of the standardized invariants $\hat{Q}$ and $\hat{R}$ for different values of $\alpha$ and from DNS. As $\alpha$ is tuned from $-1.3$ to $-0.6$, the joint PDF first extends along the left part of the Vieillefosse line, becomes symmetric for Gaussian statistics ($\alpha=-1$), and finally extends along the right part of the Vieillefosse line. This corresponds to a shift of probability from flow regions with two compressive principal strain directions to regions with two extensional principal strain directions \cite{davidson_turbulence:_2004}. For $\alpha=-0.6$, our model qualitatively captures the shape of the PDF as observed in DNS (lower right panel) and experiments \cite{meneveau_lagrangian_2011}, although the probability of velocity gradient configurations along the right part of the Vieillefosse line is underestimated. Nonetheless, since our model inherently fulfills the Betchov constraints, the mean of our model $\hat{R}$-$\hat{Q}$ PDF lies accurately at $\langle \hat{R}\rangle=\langle \hat{Q}\rangle=0$ for all values of $\alpha$.

Strain self-amplification also impacts another important aspect of the the small-scale topology: the alignment between the vorticity vector and the principal strain-rate axes. Figure \ref{fig:fig1}(c) shows the PDFs of the cosine of the angle between the vorticity vector and the three eigenvectors of the strain-rate tensor. Our model (with $\alpha=-0.6$) accurately captures the alignment of the vorticity with all three eigenvectors. In particular, it captures the well-known preferential alignment of the vorticity with the eigenvector to the intermediate eigenvalue \cite{meneveau_lagrangian_2011, ashurst_alignment_1987,xu_pirouette_2011}. The alignment strength decreases with decreasing self-amplification, and for $\alpha=-1$, when the strain self-amplification vanishes and the model statistics are Gaussian, as expected, no preferential alignment is observed.

\begin{figure}
	\includegraphics[width=\linewidth]{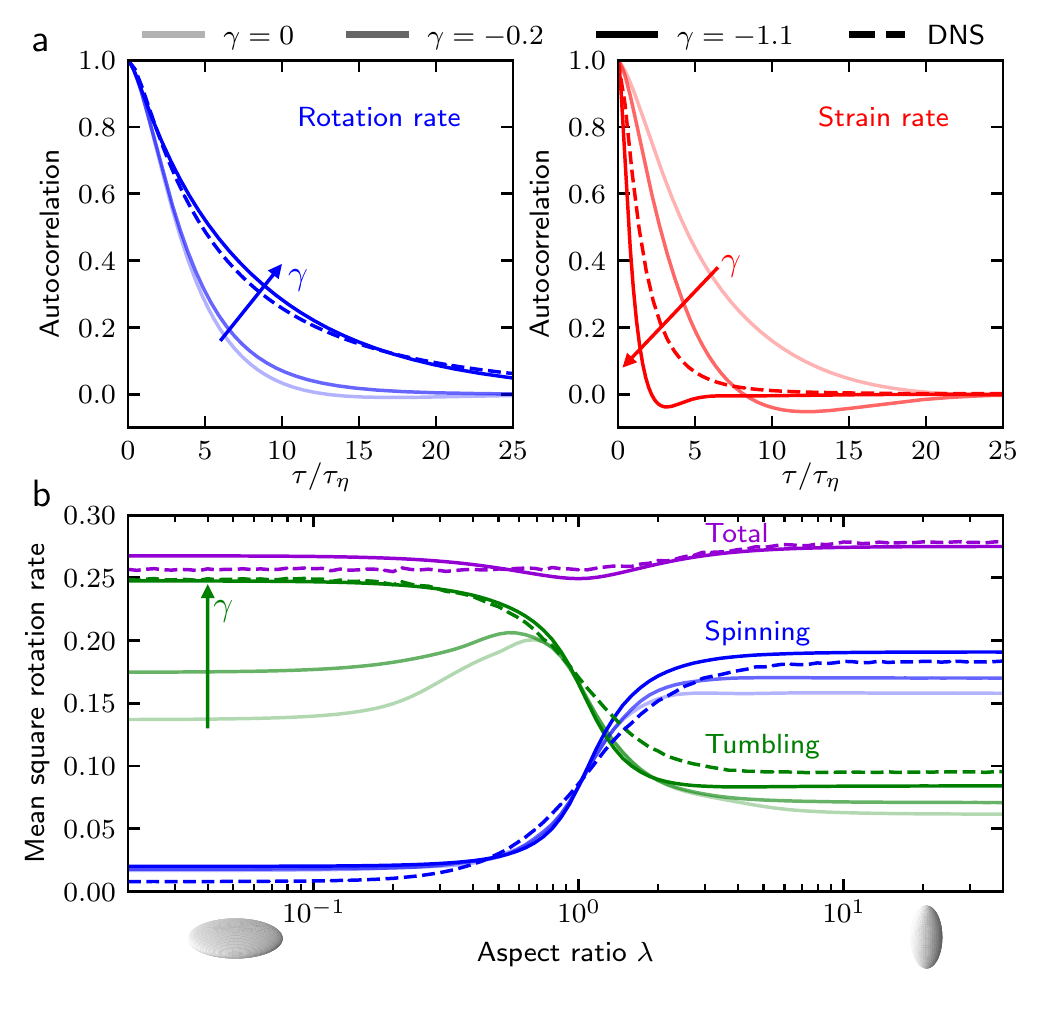}
	\caption{Controlling temporal correlations and particle rotation rates with $\gamma$. (a) Temporal auto-correlation of rotation rate (blue) and strain rate (red) for different values of $\gamma$ and DNS.
	(b) Mean square rotation rates of anisotropic particles as a function of the particles' aspect ratio as predicted by our model and DNS, non-dimensionalized by the Kolmogorov time scale. The tumbling rates (green) and spinning rates (blue) are plotted for three different values of $\gamma$ corresponding to panel (a). The total rotation rate (sum of spinning and tumbling) is plotted for $\gamma=-1.1$ in violet.}
	\label{fig:fig2}
\end{figure}

While we showed analytically that the $\gamma$-term has no effect on the single-time statistics, it determines temporal correlations of the velocity gradients. This can be rationalized from the fact that $\gamma(\bs{SW}-\bs{WS})$ essentially rotates the strain eigenframe about the axis given by the vorticity vector with a rotation rate proportional to the vorticity magnitude \cite{wilczek_pressure_2014}. This directly impacts the temporal correlation of velocity gradients and allows to precisely control them. Figure~\ref{fig:fig2}(a) compares temporal correlations of velocity gradients $\langle C_{ij}(t)C_{ij}(t+\tau)\rangle/\sqrt{\langle C_{mn}(t)^2\rangle\langle C_{pq}(t+\tau)^2\rangle}$ (implied summation) of our model to DNS results, where $C_{ij}=S_{ij}$ or $W_{ij}$. For the DNS results, the simulation was continued with $10^6$ Lagrangian tracer particles, and we collected data from the statistically stationary state. For $\gamma=-1.1$, our model matches the vorticity autocorrelation very well. Importantly, it also captures the previously observed \cite{yu_lagrangian_2010} shorter correlation time of the rate of strain compared to the rate of rotation, although differences occur in the shape of the correlation function. These results show, in particular, that the rotation of the strain eigenframe as encoded by the $\gamma$-term is responsible for a decrease of the correlation time of the strain rate and an increase of the rotation-rate correlation time.

Having established a model that captures the different temporal correlations of strain and vorticity along with the central non-Gaussian features of small-scale turbulence, we can use it to predict the tumbling and spinning rates of Lagrangian particles. To this end, we couple our model to Jeffery's equation \cite{jeffery_motion_1922}, which describes the orientational dynamics of axisymmetric ellipsoidal particles (implied summation),
\begin{equation}
\label{eq:jeffrys}
\frac{d}{dt}p_i=W_{ij}p_j+\frac{\lambda^2-1}{\lambda^2+1}(S_{ij}p_j-p_ip_kS_{kl}p_l) \, .
\end{equation}
Here, the unit vector $\bs p$ denotes a particle's symmetry axis, and $\lambda$ is the particle's aspect ratio, i.e., the ratio of the length along the symmetry axis to the length perpendicular to it. The rotation of an axisymmetric particle can be decomposed into spinning (rotation around the symmetry axis) and tumbling (rotations around an axes perpendicular to the symmetry axis) \cite{voth_anisotropic_2017}, with the squared spinning rate $\left(\frac{1}{2}\bs{\omega}\cdot\bs{p}\right)^2$, where $\bs{\omega}$ is the vorticity, and the squared tumbling rate $\dot{p_i}\dot{p_i}$. In Fig.~\ref{fig:fig2}(b) the nondimensionalized mean square tumbling and spinning rates as predicted by our model are shown for different values of $\gamma$ as a function of the particles' aspect ratio. Figure \ref{fig:fig2}(b) shows that especially the tumbling rates of disklike particles increase when the temporal correlations are modified by increasing the magnitude of the coefficient $\gamma$. When our model exhibits the most realistic correlation times, i.e., for $\gamma=-1.1$, the particle rotation rates predicted by our model agree very well with the ones observed in our DNS and literature \cite{voth_anisotropic_2017,byron_shape-dependence_2015} for the full range of particle shapes. In particular, our model predicts the high tumbling rates of disklike particles observed in DNS and experiments. The comparison of the results for different values of $\gamma$ in Figs.~\ref{fig:fig2}(a) and \ref{fig:fig2}(b) indicates that the realistic auto-correlation times of our model are crucial for an accurate prediction of tumbling rates of suspended particles. 

In summary, we have analyzed the dynamics and statistics of velocity gradients in turbulence in the framework of a minimal, physically consistent reduced-order model. Our combined analytical and computational analysis showed that strain self-amplification controls the non-Gaussianity as well as the small-scale topology of the velocity gradient dynamics, and identified the rotation of the strain eigenvectors by the vorticity as the major factor in determining the temporal correlations of velocity gradients. As a result, we obtained a reduced-order model for the small scales of turbulence that captures the different correlation times of strain and vorticity in turbulence. We showed that the reduced-order model can be used to accurately predict the orientational statistics of suspended anisotropic particles, enabling a host of modeling applications for complex particulate flows.

Based on tensor function representation theory and the systematic implementation of physical constraints, our closure approach explicitly uncovers the general tensorial structure of the unclosed terms, which also provides a firm foundation for future advancements. For example, we expect that the inclusion of higher-order terms and coefficient functions that depend on velocity gradient tensor invariants will lead to further quantitative improvement. Machine learning approaches \cite{tian_physics_2019} could turn out to be instrumental in achieving such improved parameterizations.

\begin{acknowledgments}

We thank Jos\'e-Agust\'in Arguedas-Leiva, Tobias B\"atge, Cristian Lalescu, and B\'erenger Bramas for scientific computing support and the provision of the DNS data. Valuable feedback on the manuscript from Lukas Bentkamp and Maurizio Carbone is gratefully acknowledged. The authors gratefully acknowledge the Gauss Centre for Supercomputing e.V. for funding this project by providing computing time on the GCS Supercomputer SuperMUC at Leibniz Supercomputing Centre. Computational resources from the Max Planck Computing and Data Facility and support by the Max Planck Society are gratefully acknowledged.
\end{acknowledgments}

\bibliography{references}

\input{supplements.tex}
\end{document}

%% file: supplements.tex
\renewcommand{\thefigure}{S\arabic{figure}}
\renewcommand{\theequation}{S\arabic{equation}}
\stepcounter{myequation}
\stepcounter{myfigure}
\onecolumngrid
\pagebreak
\newpage 

\begin{center}
	\textbf{\large Supplemental Material: Capturing Velocity Gradients and Particle Rotation Rates in Turbulence}\\[.2cm]
	Leonhard A. Leppin$^{1,2,3}$ and Michael Wilczek$^{1,2}$\\[.1cm]
	{\itshape \mbox{${}^1$Max Planck Institute for Dynamics and Self-Organization (MPI DS),
		Am Fa\ss berg 17, 37077 G\"ottingen, Germany}\\
		\mbox{${}^2$Faculty of Physics, Georg-August-Universit\"at G\"ottingen, Friedrich-Hund-Platz 1, 37077 G\"ottingen, Germany}\\
		${}^{3}$Max Planck Institute for Plasma Physics, Boltzmannstra\ss e 2, 85748 Garching, Germany
	}\\[1cm]
\end{center}

\section{Structure of unclosed terms based on tensor function representation theory}
\label{sec:sm1}
To obtain a closed statistical evolution equation for the velocity gradient tensor (Eq.~(1) in the main text), the conditional mean pressure Hessian and the conditional mean viscous Laplacian terms need to be specified. Both terms are tensor-valued functions of the velocity gradient tensor, whose structure is constrained by statistical isotropy. For a second-order tensorial function $\bs M$ which depends on the velocity gradient tensor $\bs A$, statistical isotropy implies
\begin{equation}
  \bs{M}(\bs{Q}\bs{A}\bs{Q}^T)=\bs{Q}\bs{M}(\bs{A})\bs{Q}^T
\end{equation} 
where $\bs{Q}$ is an arbitrary orthogonal matrix. Additional constraints arise when $\bs M$ is symmetric or antisymmetric and depends on symmetric or antisymmetric tensorial arguments. The general structure of the tensorial function can then be determined using tensor function representation theory \cite{zheng_theory_1994,wang_new_1970,smith_isotropic_1971,boehler_irreducible_1977}, which expresses the function as a linear combination of tensorial terms (also called generators or form-invariants) with scalar coefficient functions that may depend on scalar tensor invariants of $\bs{A}$. In the mathematical literature representations for two kinds of functions exist: For polynomial \cite{spencer_theory_1958,spencer_further_1959,pope_more_1975} and for general tensor functions \cite{zheng_theory_1994,wang_new_1970,smith_isotropic_1971,boehler_irreducible_1977}. For the closure, we use representations for general tensor functions, as these do not require assumptions about the functional form and also contain in general fewer terms than the representations of polynomial functions \cite{boehler_applications_1987,zheng_theory_1994}. Based on this, the conditional traceless and symmetric pressure Hessian takes the form:
\begin{equation}
\langle {\widetilde{\bs{H}}}|\bs{A}\rangle=\sum_{n=1}^{7}b^{(n)}\bs{B}^{(n)}
\end{equation}

where the $\bs{B}^{(n)}$ are: 
\begin{align*}
&\bs{B}^{(1)}=\bs{S} & &\bs{B}^{(2)}=\bs{S}^2-\frac{1}{3}\Tr(\bs{S}^2)\bs{I}& &\bs{B}^{(3)}=\bs{W}^2-\frac{1}{3}\Tr(\bs{W}^2)\bs{I} \\
&\bs{B}^{(4)}=\bs{S}\bs{W}-\bs{W}\bs{S}& &\bs{B}^{(5)}=\bs{S}\bs{W}^2+\bs{W}^2\bs{S}-\frac{2}{3}\Tr(\bs{SW}^2)\bs{I}& &\bs{B}^{(6)}=\bs{S}^2\bs{W}-\bs{W}\bs{S}^2 \\
&\bs{B}^{(7)}=\bs{W}\bs{S}\bs{W}^2-\bs{W}^2\bs{S}\bs{W}
\end{align*}
and the coefficients $b^{(n)}$ are functions of the scalar invariants  $\Tr(\bs{S}^2)$, $\Tr(\bs{S}^3)$, $\Tr(\bs{W}^2)$, $\Tr(\bs{S}\bs{W}^2)$, $\Tr(\bs{S}^2\bs{W}^2)$, and $\Tr(\bs{S}^2\bs{W}^2\bs{S}\bs{W})$. Note that there are six independent scalar invariants of $\bs{S}$ and $\bs{W}$ (and hence $\bs{A}$), i.e.~one independent scalar invariant more than typically discussed in the turbulence literature \cite{pope_turbulent_2000, davidson_turbulence:_2004,meneveau_lagrangian_2011}. This difference is rooted in the fact that the number of invariants in an irreducible functional basis of a tensor is not necessarily equal to the number of independent entries of the tensor, which is typically considered. Furthermore, an irreducible functional basis is not necessarily minimal, meaning that a different representation with fewer elements might exist \cite{pennisi_irreducibility_1987,zheng_theory_1994}. The irreducibility of the functional basis presented here has been explicitly proven in \cite{pennisi_irreducibility_1987}.

Similar to the conditional pressure Hessian, one can construct the most general tensor function representation for the viscous Laplacian term. Since the function representations only hold for symmetric or antisymmetric tensor functions, one first has to decompose the Laplacian of $\bs{A}$ into the Laplacian of its symmetric part $\bs{S}$ and antisymmetric part $\bs{W}$. We find:
\begin{equation}
\langle \nu\Delta \bs{A}|\bs{A}\rangle
=\langle \nu \Delta \bs{S}|\bs{A} \rangle+\langle \nu \Delta \bs{W}|\bs{A} \rangle =\sum_{n=1}^{7}c^{(n)}\bs{B}^{(n)} + \sum_{n=1}^{3}d^{(n)}\bs{D}^{(n)}
\end{equation}
where the tensors for the representation of the antisymmetric part are 
$\bs{D}^{(1)}=\bs{W}$, $\bs{D}^{(2)}=\bs{S}\bs{W}+\bs{W}\bs{S}$,
$\bs{D}^{(3)}=\bs{S}\bs{W}^2-\bs{W}^2\bs{S}$.
The $c^{(n)}$ and $d^{(n)}$ are functions of the same scalar invariants as the $b^{(n)}$.

\section{Derivation of the adaptive coefficients}
\label{sec:sm2}
\begin{figure}
	\includegraphics[width=.8\textwidth]{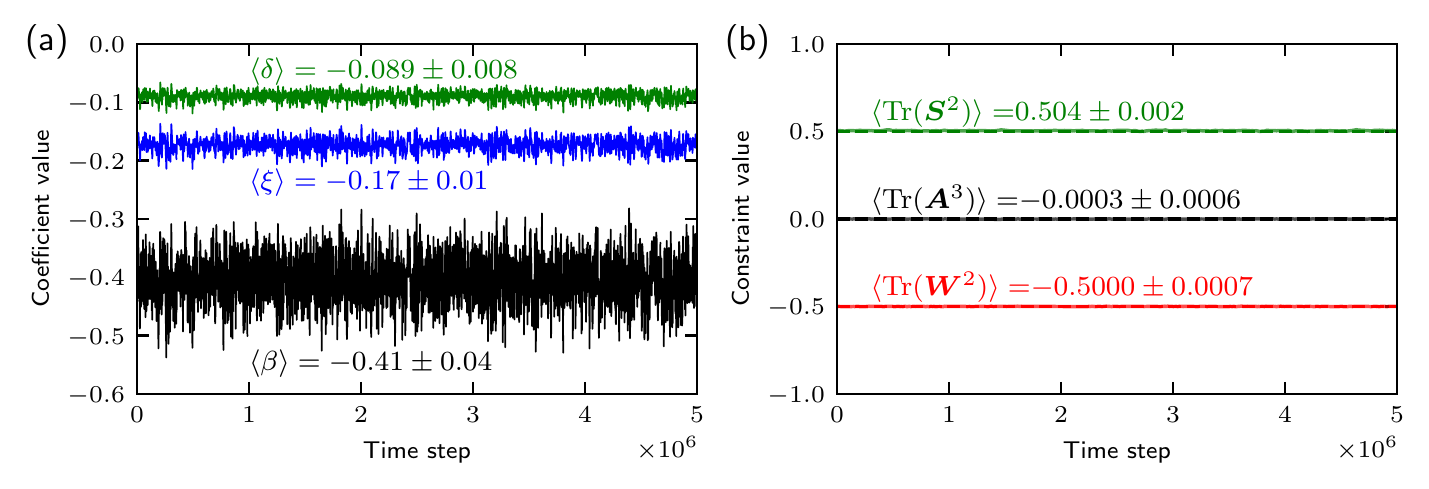}
	\caption{(a) Values of the adaptive coefficients as a function of time for a simulation with $\alpha=-0.6$, $\sigma=0.08$, $\gamma=-1.1$. (b) Empirical values of the imposed constraints for the same simulation as in (a). Dashed: imposed values, solid: actual values during simulation. The error margins are determined as the standard deviation from the mean.}
	\label{fig:app1}
\end{figure}

The closure for the velocity gradient model, as given by Eqs.~(2) and (3) in the main text, depends on five coefficients. The number of independent coefficients can be reduced by nondimensionalizing the equation of motion (Eq.~(1) in the main text) by the Kolmogorov time scale $\tau_\eta$, which implies $\langle \mathrm{Tr}(\bs S^2) \rangle = 1/2$. Here and in the following, $\langle \cdot\rangle$ denotes an ensemble average. The Betchov constraints for homogeneous turbulence, $\langle \mathrm{Tr}(\bs A^2) \rangle = 0$ and $\langle \mathrm{Tr}(\bs A^3) \rangle = 0$, can be used to further reduce the number of parameters to a total of two independent parameters, $\alpha$ and $\gamma$. For the three model coefficients fixed in this way, analytical expressions can be found. They are constructed by deriving differential equations from the model SDE for each constrained quantity ($\langle\Tr(\bs{S}^2)\rangle$, $\langle\Tr(\bs{W}^2)\rangle$, $\langle\Tr(\bs{A}^3)\rangle$) using It\^o's formula. After averaging, one finds, for example, the following equation for $\langle\Tr(\bs{S}^2)\rangle$:
\begin{equation}
d\langle S_{ab}S_{ba}\rangle
=\left\langle\frac{\partial S_{ab}S_{ba}}{\partial A_{ij}}(N_{ij}+L_{ij})\right\rangle dt +\frac{1}{2}\left\langle\frac{\partial^2 S_{ab}S_{ba}}{\partial A_{ik}\partial A_{jl}}\right\rangle Q_{ijkl}(\bs{0})dt
\end{equation}
and accordingly for the other constraints. Here, $N_{ij}$ and $L_{ij}$ are the nonlinear and linear terms of the FPE (Eqs.~(5) and (6) in the main text), respectively, and $Q_{ijkl}(\bs{0})$ denotes the forcing covariance, detailed in section \ref{sec:forcing}. By carrying out the derivatives and inserting $N_{ij}$ and $L_{ij}$, the averages can be evaluated. As for constant constraints the left-hand sides vanish, we can solve for three of the parameters and obtain
\begin{equation}
\xi =2\langle\epsilon\Tr(\bs{W}^2)\rangle-\frac{15}{2}\sigma^2-4\langle\Tr(\bs{SW}^2)\rangle
\end{equation}

\begin{equation}
\delta=2\langle\Tr(\bs{SW}^2)\rangle(3\alpha-\beta)+2\langle\epsilon\Tr(\bs{A}^2)\rangle
\end{equation}

\begin{equation}
\beta=\frac{\langle\Tr(\bs{A}^2\widetilde{\bs{A}^2})\rangle+\alpha\left(\langle\Tr(\bs{A}^2\widetilde{\bs{S}^2})\rangle+6\langle\Tr(\bs{SW}^2)\rangle\langle\Tr(\bs{A}^2\bs{S})\rangle\right) +2\langle\epsilon\Tr(\bs{A}^2)\rangle\langle\Tr(\bs{A}^2\bs{S})\rangle-\langle\epsilon\Tr(\bs{A}^3)\rangle}{2\langle\Tr(\bs{SW}^2)\rangle\langle\Tr(\bs{A}^2\bs{S})\rangle-\langle\Tr(\bs{A}^2\widetilde{\bs{W}^2})\rangle}
\end{equation}
where $\epsilon$ is the auxiliary damping term, discussed in the main text and in section \ref{sec:dampingterm}. These expressions are evaluated in every simulation time step to update the coefficient values. Fig.~\ref{fig:app1}(a) shows an example of the parameter values which are dynamically obtained throughout a simulation. Fig.~\ref{fig:app1}(b) demonstrates that the constraints are indeed fulfilled to very good precision throughout the simulation.

\section{Impact of the forcing amplitude}
\label{sec:forcing}
\begin{figure}
	\includegraphics[width=\textwidth]{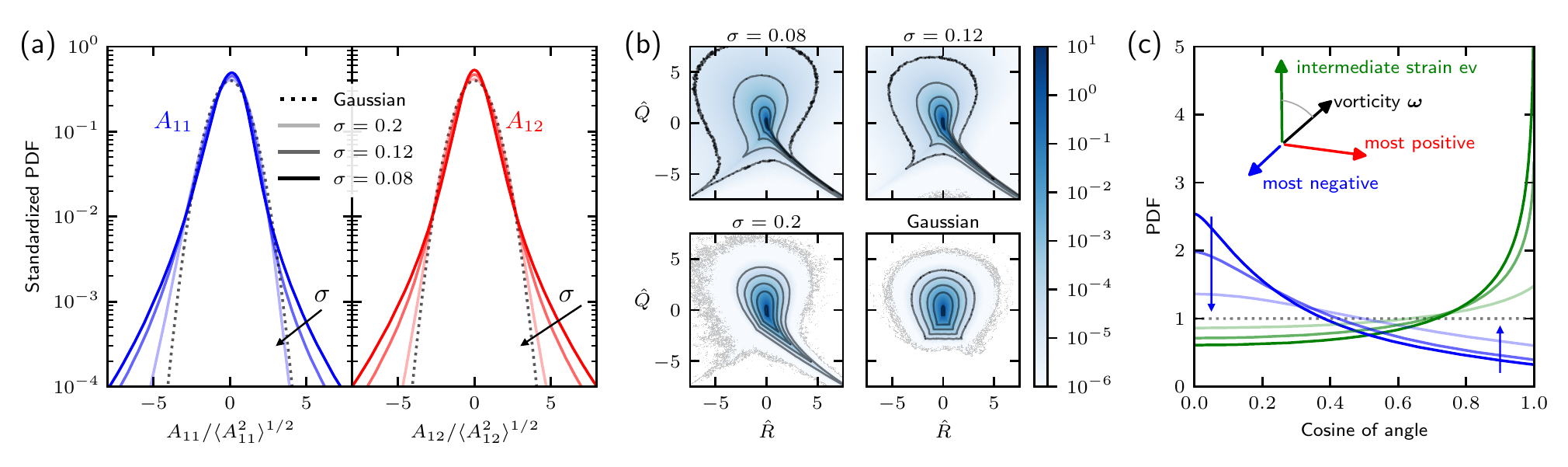}
	\caption{(a) Standardized PDFs of the longitudinal $A_{11}$ and transverse $A_{12}$ velocity gradient components for different forcing amplitudes $\sigma$. (b) Joint PDF of the standardized isotropic invariants $\hat{R}=-\frac{1}{3}\frac{\Tr(\bs{A}^3)}{\langle\Tr(\bs{S}^2) \rangle^{3/2}}$ and $\hat{Q}=-\frac{1}{2}\frac{\Tr(\bs{A}^2)}{\langle\Tr(\bs{S}^2)\rangle}$ for different values of $\sigma$. (c) Alignment of principal strain axes and vorticity. PDFs of the cosine of the angle between the vorticity vector and the most negative, intermediate eigenvectors of the strain-rate tensor for the same values of $\sigma$  as in panel (a). The alignment of the vorticity with the most positive eigenvector is very weak for all values of $\sigma$. It is therefore omitted to improve the visual clarity.}
	\label{fig:app2}
\end{figure}

The forcing term $dF_{ij}=\sigma D_{ijab}dW_{ab}$ is based on a tensorial Gaussian white noise $dW_{ab}$ with statistically independent components ($\langle dW_{ab} \rangle=0$, $\langle dW_{ac}dW_{bd}\rangle=\delta_{ab}\delta_{cd}dt$), and has a covariance $\langle dF_{ik}dF_{jl}\rangle=Q_{ijkl}(\bs{0})dt$ with
\begin{equation}
\label{eq:forcingcov}
Q_{ijkl}(\bs{0})=\sigma^2D_{ikab}D_{jlab}=\frac{\sigma^2}{2}(4\delta_{ij}\delta_{kl}-\delta_{ik}\delta_{jl}-\delta_{il}\delta_{jk})
\end{equation}
which is designed to be consistent with isotropy, homogeneity and tracelessness of $\bs{A}$ \cite{chevillard_modeling_2008,wilczek_pressure_2014} in the model SDE (Eq.~(1) in the main text). It can be motivated from a stochastic Gaussian force term in the Navier-Stokes equation \cite{wilczek_pressure_2014} and ensures stationary statistics by counteracting dissipative terms of the model. For small values of $\sigma$ ($\sigma\approx0.05$), the deterministic dynamics are only little perturbed, leading to more excursive trajectories. Larger values of $\sigma$ ($\sigma\approx0.2$) regularize the dynamics by increasing the influence of the Gaussian forcing. For large forcing amplitudes $\sigma$, the forcing dominates over the deterministic dynamics of the model, leading to Gaussian statistics. Therefore, as $\sigma$ increases, the tails of the component PDFs become less pronounced and the preferential alignments become weaker (see Fig.~\ref{fig:app2}(a) and (c)). Furthermore, the probability density along the Vieillefosse line in the lower left quadrant of the $R$--$Q$ plane decreases (see Fig.~\ref{fig:app2}(b)). 
Based on a parameter scan of the forcing amplitude $\sigma$ and comparison with direct numerical simulations, we set $\sigma=0.08$.

\section{Analysis of the Fokker-Planck equation}

\subsection{Gaussian solution for $\alpha=-1$}
\label{sec:sm41}
The strain self-amplification, controlled by the parameter $\alpha$, determines the degree of Gaussianity of the modeled velocity gradient statistics. In fact, for $\alpha=-1$, the Gaussian velocity gradient PDF
\begin{equation}\label{eq:gaussiansolution}
  g(\bs{A})=\frac{225\sqrt{5}}{16\pi^4\langle\Tr(\bs{S}^2)\rangle^4}\exp{\left(-\frac{1}{2}A_{ik}R_{ijkl}^{-1}A_{jl}\right)}\delta(\Tr(\bs{A}))
\end{equation}
is an exact solution to the Fokker-Planck equation (Eq.~(4) in the main text), which is consistent with the Betchov constraints.  Here, $R^{-1}_{ijkl}=\frac{1}{\langle \Tr(\bs{S}^2)\rangle}[4\delta_{ij}\delta_{kl}+\delta_{il}\delta_{jk}]$ is the pseudo-inverse of the velocity gradient covariance tensor $R_{ijkl} = \langle A_{ik} A_{jl} \rangle$. The delta function stems from the incompressibility condition of the velocity field, $\Tr(\bs{A})=0$.

To see that \eqref{eq:gaussiansolution} solves the Fokker-Planck equation, we consider the nonlinear drift term. To this end, it is important to note that $\frac{\partial N_{ij}}{\partial A_{ij}}=0$. We then obtain 
\begin{equation}
\begin{split}
  \frac{\partial}{\partial A_{ij}}\left[ N_{ij}g(\bs{A}) \right] &=N_{ij}\frac{\partial}{\partial A_{ij}}g(\bs{A}) \\
  &=N_{ij}\frac{1}{\langle\Tr(\bs{S}^2)\rangle}[-5S_{ij}-3W_{ij}]g(\bs{A})\\
  &=\frac{1}{\langle\Tr(\bs{S}^2)\rangle}[5(1+\alpha)\Tr(\bs{S}^3)-\Tr(\bs{SW}^2)+5\beta \Tr(\bs{SW}^2)]g(\bs{A}) \, .
\end{split}
\end{equation}
This expression vanishes for $\alpha=-1$ and $\beta = 1/5$, which implies that all nonlinear terms can cancel when the Gaussian PDF is inserted. Note that for $\alpha=-1$, the Betchov constraints already require $\beta = 1/5$, such that effectively only $\alpha=-1$ needs to be fixed. The Gaussian PDF \eqref{eq:gaussiansolution} solves the remaining FPE if $\delta=0$ and $\xi=-15\sigma^2/[4\langle \Tr(\bs{S}^2)\rangle]$, which is consistent with the Betchov constraints.

\subsection{Independence of the single-time statistics of $\gamma$}
\label{sec:sm42}
Here we show that the single-time statistics is independent of $\gamma$ under the assumption of statistical isotropy. This allows us to control the temporal correlations in our model through $\gamma$ without changing the single-time statistics. For isotropic statistics, the velocity gradient PDF is a scalar function of the six isotropic invariants of $\bs A$ \cite{zheng_theory_1994}:
\begin{equation}
\label{eq:f(A)}
  f(\bs{A}) = F\left(\Tr(\bs{S}^2), \Tr(\bs{S}^3), \Tr(\bs{W}^2), \Tr(\bs{SW}^2), \Tr(\bs{S}^2\bs{W}^2), \Tr(\bs{S}^2\bs{W}^2\bs{SW})\right) \, .
\end{equation}
For the $\gamma$-term in the Fokker-Planck equation (Eq.~(4) in the main text), we therefore obtain
\begin{equation}
\begin{split}
\label{eq:gammaterm}
\frac{\partial}{\partial A_{ij}}\left[\gamma(S_{ik}W_{kj}-W_{ik}S_{kj})f(\bs{A})\right] &= \gamma(S_{ik}W_{kj}-W_{ik}S_{kj})\frac{\partial}{\partial A_{ij}} f(\bs{A})  \\
&= \gamma(S_{ik}W_{kj}-W_{ik}S_{kj}) \sum_{n=1}^6 \frac{\partial X_n}{\partial A_{ij}} \frac{\partial F}{\partial X_n}
\end{split}
\end{equation}
where the $X_n$ are the six scalar invariants given in Eq.~\eqref{eq:f(A)}. One can now straightforwardly compute that the tensor contraction $\gamma(S_{ik}W_{kj}-W_{ik}S_{kj}) \frac{\partial X_n}{\partial A_{ij}} = 0$ term by term, which shows that the $\gamma$-term vanishes in the Fokker-Planck equation.

\section{Simulation details}
\label{sec:sm5}
The integration of the SDE (Eq.~(1) in the main text) has been implemented in Python using the Euler-Maruyama method with a time step size $\Delta t = 0.0002$. For all simulations shown here, we have integrated an ensemble of $10^5$ Gaussian initial conditions for $5\times10^6$ time steps, which corresponds to $1000\tau_\eta$ in physical time, after an initial transient of $100\tau_\eta$.  

Furthermore, an integration of Jeffery's equation (Eq.~(7) in the main text) has been implemented to obtain the rotation rates of axisymmetric ellipsoidal Lagrangian particles. Initially, $10^4$ randomly oriented symmetry axes of the particles were created for each particle aspect ratio and then evolved with Jeffery's equation using the Euler method, where $\bs{S}$ and $\bs{W}$ were taken from our model simulation. The simulation of the particles was done for logarithmically spaced aspect ratios ranging from 0.02 to 40.

\section{Rogue trajectories and nonlinear damping term}
\label{sec:dampingterm}

Initial simulations of the SDE (Eq.~(1) in the main text) revealed the occurrence of rare rogue trajectories exploring far-out regions of the phase space. To understand their occurrence in our model, it is particularly instructive to consider the unforced ($d\bs{F}=\bs 0$) and undamped ($\delta=0$, $\xi=0$) case, and to decompose the model into the rate-of-strain and rate-of-rotation dynamics. Let us focus on a particular set of initial conditions with $\bs W(0) = \bs 0$. Because of $d \bs W/d t = \bs 0$, the set of differential equations then reduces to
$
d \bs S/dt = -(1+\alpha) \widetilde{\bs{S}^2} \, .
$
This equation has a finite-time singularity when $\alpha \neq -1$, similar to the one in the Restricted Euler Model \cite{vieillefosse_internal_1984,vieillefosse_local_1982}. This shows that our SDE in the undamped and unforced case includes at least one sub-manifold of initial conditions which leads to a finite-time singularity.

Including a linear damping term, as, for example, also done in the framework of linear diffusion models \citep{martin_1998}, stabilizes some regions of the phase space. However, large velocity gradients still diverge. The stochastic forcing, however, has a profound impact. Because trajectories are randomly kicked off the unstable sub-manifolds, which are presumably of zero measure, the overall dynamics is stabilized. Still, trajectories close to these unstable manifolds can explore far-out regions of the phase space. These are the rogue trajectories.

In our model with stochastic forcing and linear damping only, we find that the rogue trajectories are over-represented compared to actual Navier-Stokes turbulence. The frequent occurrence of the rogue trajectories can furthermore destabilize our numerical scheme which is based on adaptive coefficients that depend on ensemble averages and thereby couple the trajectories with each other. The impact of the rogue trajectories can be mitigated by the inclusion of a nonlinear damping term as explained in the following. In a nonlinear system, singularities can be effectively prevented if the highest-order nonlinearity has a damping influence. Because the Restricted Euler terms and the pressure Hessian closure are quadratic nonlinearities, they can be stabilized by a damping nonlinearity of cubic order or higher. This is achieved by including the damping factor as documented here and in the main text. We chose the comparably high-order nonlinearity $\epsilon = -10^{-8}[(\Tr(\bs{W}^2)+1/2)^4+(\Tr(\bs{S}^2)-1/2)^4]$ to leave the dynamics in the central region of phase space untouched and only prevent potentially diverging rogue trajectories. $\epsilon$ is fully compatible with the model approach based on tensor function representation theory, which explicitly includes scalar prefactors that depend on velocity gradient tensor invariants.

\begin{figure}
	\centering
	\includegraphics[width=0.4\textwidth]{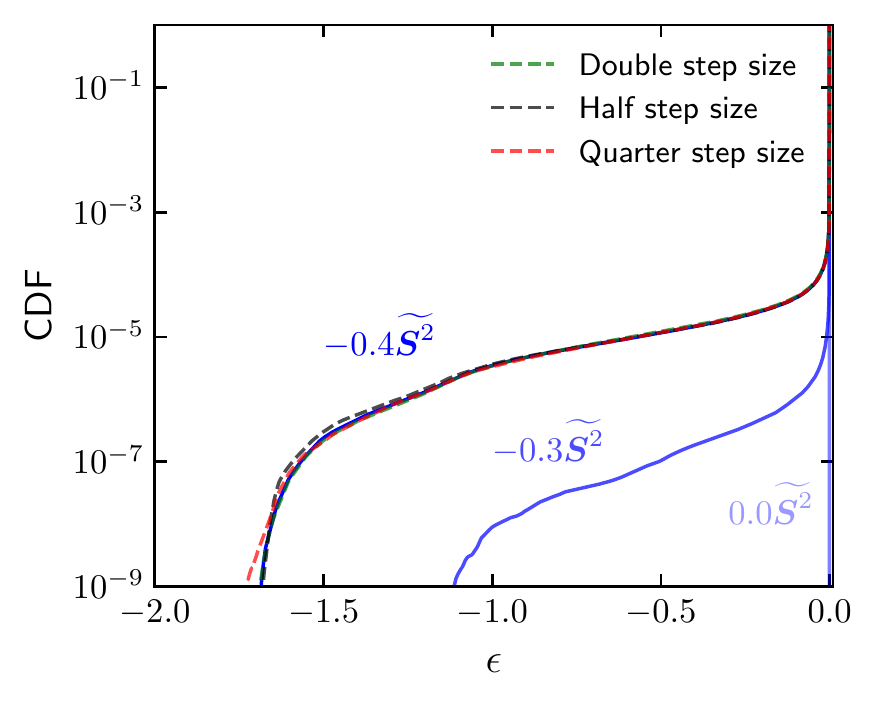}
	\caption{Cumulative distribution function of the auxiliary nonlinear term $\epsilon$ for three different values of $\alpha$ (left  to right: $-0.6$, $-0.7$,$-1$), i.e.~different prefactors of $\widetilde{\bs{S}^2}$. As the strain self-amplification becomes more important, larger magnitudes of $\epsilon$ are observed. For the most relevant parameter choice discussed in the main text, $\alpha=-0.6$, i.e.~$-0.4\widetilde{\bs{S}^2}$, the distribution is shown for three additional time step sizes, demonstrating convergence of our numerical scheme.
	}
	\label{fig_appendix3}
\end{figure}

One way to characterize the occurrence of rogue trajectories in our simulations is to analyze the statistics of the nonlinear damping term because this term will only become relevant if trajectories are visiting far-out regions of the phase space. In particular, we expect that the damping factor increases with increasing influence of the strain self-amplification term $\widetilde{\bs{S}^2}$, which - as explained above - causes the finite-time singularity of the undamped system. In Fig.~\ref{fig_appendix3} the cumulative distribution function of $\epsilon$ for different values of $\alpha$ is shown. Indeed, as the prefactor $-(1+\alpha)$ deviates from zero, larger absolute values of $\epsilon$ are observed. This shows how strain self-amplification gives rise to more extreme trajectories, which then are damped in far-out regions of the phase space through the $\epsilon$-term. In simulations of the analytically accessible Gaussian case with $\alpha = -1$, in which the strain self-amplification vanishes, no rogue trajectories are observed, as expected. As a result, $\epsilon$ is sharply localized at zero.

To demonstrate numerical convergence of the case $\alpha=-0.6$, we performed additional simulations with significantly smaller time steps (Fig.~\ref{fig_appendix3}). As we find convergent results, we conclude that our time stepping scheme used for the integration of the SDE accurately represents the dynamics of the system.
\\